\documentclass[aps,prl,english,superscriptaddress,twocolumn]{revtex4}

\usepackage{amsmath,amssymb,amsfonts}
\usepackage{CJK}
\usepackage{bm}
\usepackage{tikz}
\usepackage{babel}
\usepackage{blindtext}

\def \Z {\mathbb{Z}}

\def \i {\mathrm{i}}

\usepackage[colorlinks=true, letterpaper=true, pdfstartview=FitV, linkcolor=blue, citecolor=blue, urlcolor=blue]{hyperref}

\begin{document}

\title{Topological second-order spin-$3/2$ liquids with hinge Fermi arcs}

\author{Y. X. Zhao}
\email[]{zhaoyx@nju.edu.cn}
\affiliation{National Laboratory of Solid State Microstructures and Department of Physics, Nanjing University, Nanjing 210093, China}
\affiliation{Collaborative Innovation Center of Advanced Microstructures, Nanjing University, Nanjing 210093, China}

\author{Y. Lu}
\email[]{y.lu@thphys.uni-heidelberg.de}
\affiliation{Institute for Theoretical Physics, Heidelberg University, Philosophenweg 19, 69120 Heidelberg, Germany}

\author{Shengyuan A. Yang}
\address{Research Laboratory for Quantum Materials, Singapore University of Technology and Design, Singapore 487372, Singapore}


\begin{abstract}
	
We present an exactly solvable spin-3/2 model defined on a pentacoordinated three-dimensional graphite lattice, which realizes a novel quantum spin liquid with second-order topology. The exact solutions are described by Majorana fermions coupled to a background $\mathbb{Z}_2$ gauge field, whose ground-state flux configuration gives rise to an emergent off-centered spacetime inversion symmetry. The symmetry protects topologically nontrivial band structures for the Majorana fermions, particularly including nodal-line semimetal phases with twofold topological charges: the second Stiefel-Whitney number and the quantized Berry phase.
The former leads to rich topological phenomena on the system boundaries. There are two nodal-line semimetal phases hosting hinge Fermi arcs located on different hinges, and they are separated by a critical Dirac semimetal state with surface helical Fermi arcs. In addition, we show that rich symmetry/topology can be explored in our model by simply varying the lattice or interaction arrangement. As an example, we discuss how to achieve a topological gapped phase with surface Dirac points.

\end{abstract}

\maketitle

{\color{blue}\textit{Introduction.}}
Topological phases of quantum matter have been one of the most active fields in the last decade. For noninteracting systems (also including the interacting ones which allow a mean field description), we already have the well-developed topological band theory, solidly grounded by the $K$-theory, for classifying and characterizing their topological phases~\cite{Schnyder-classification,Kitaev-Classification,Schnyder-RMP,Zhao-Classification,Zhao-PRB,Zhao-Schnyder-Wang-PT,PhysRevX.7.041069}. Meanwhile, for strongly interacting systems, the task is much more challenging, and typically requires
complicated treatments, such as topological field theory, tensor category, and projective group representations.

There is, however, a remarkable exception --- the Kitaev model~\cite{Kitaev-anyons}, where a strongly interacting system permits a characterization via the topological band theory. The Kitaev model and its variants are spin-1/2 models defined on tricoordianted lattices. The model has attracted tremendous interest, because it offers the first example for which a quantum spin liquid (QSL) can be unambiguously identified in the ground state, and furthermore, this QSL can possess nontrivial topologies~\cite{Kitaev-anyons,YaoHong-Kivelson,PhysRevB.76.180404,YaoHong-SU2,PhysRevB.93.085101}. By varying the model parameters and the lattice geometry, a number of topological features have been found in the Kitaev QSL, ranging from (Majorana) Fermi surfaces~\cite{PhysRevB.89.235102,PhysRevLett.115.177205}, nodal lines~\cite{PhysRevB.79.024426,PhysRevLett.114.116803}, Weyl points~\cite{PhysRevLett.114.157202}, to gapped states with topological edge~\cite{Kitaev-anyons} or corner modes~\cite{PhysRevB.98.054432}.


In this Letter, we present a new topological QSL beyond the Kitaev-model paradigm. We construct an \emph{exactly solvable} model for \emph{spin-3/2} on a \emph{pentacoordinated} lattice, e.g., the three-dimensional (3D) graphite lattice, which consists of bond-dependent quartic interactions between neighboring spins. Analogous to the Kitaev model, an exact solution of our model can be obtained by fractionalizing the spin-3/2 into a set of Majorana fermions and mapping the strongly interacting system to a system of noninteracting Majorana fermions hopping in a background $\mathbb{Z}_2$ gauge field. We show that the obtained QSL ground state is gapless for the Majorana spinors and gapped for gauge vortices. Interestingly, the ground-state gauge configuration leads to an emergent twofold screw rotational symmetry. Thereby, novel \emph{second-order} topological Majorana ``semimetals" are realized, protected by an emergent off-centered spacetime inversion ($\mathcal{P}T$) symmetry.
The two generic phases possess a pair of nodal loops in the bulk and topological Fermi arcs on a single pair of diagonal or off-diagonal hinges~\cite{wang2020boundary}. The critical point between the two phases is a 3D Dirac semimetal with a pair of bulk Dirac points and helical surface Fermi arcs. We show that the nodal loops and the Dirac points both feature a nontrivial $\mathbb{Z}_2$ topological charge given by the second Stiefel-Whitney number $\nu_\text{2D}$~\cite{Zhao-Schnyder-Wang-PT,Zhao-PT-Dirac,PhysRevB.96.155105,BJYang-Nodal-Line,B-J-Yang19APRPRX}, and the loops have an extra $\mathbb{Z}_2$ charge $\nu_\text{1D}$ given by the quantized Berry phase.  

The significance of our work is at least threefold: (1) It offers the first exactly solvable QSL model beyond spin-1/2; (2) it reveals new topological strongly interacting states not known before, particularly the second-order nodal-line QSL phase (we also discuss how our model can be extended to achieve other interesting topologies); (3) it demonstrates the essential importance of $\Z_2$ gauge configuration in the formation of time-reversal-invariant topological phases.

{\color{blue}\textit{Spin-3/2 and $Cl_5$.}} Before constructing our model, we first present a connection between spin-3/2 and the Clifford algebra $Cl_5$, which parallels that between spin-1/2 and $Cl_3$~\cite{Spin-Clifford}.

The five Dirac matrices $\hat{\Gamma}^a$ ($a=1,2,\cdots,5$) form a complete set of generators for $Cl_5$. They can be expressed
as quadratic forms of the three spin-$3/2$ operators $\hat{S}_\alpha$ ($\alpha=x,y,z$), as the following:
\begin{equation}\label{Gamma-operators-1}
\begin{split}
\hat{\Gamma}^1=\frac{1}{\sqrt{3}}(\hat{S}_{y}\hat{S}_{z}+\hat{S}_{z}\hat{S}_{y}),~~&\hat{\Gamma}^2=\frac{1}{\sqrt{3}}(\hat{S}_{z}\hat{S}_{x}+\hat{S}_{x}\hat{S}_{z}),\\\hat{\Gamma}^3 =\frac{1}{\sqrt{3}}(\hat{S}_{x}&\hat{S}_{y}+\hat{S}_{y}\hat{S}_{x}),
\end{split}
\end{equation}
and
\begin{equation}\label{Gamma-operators-2}
\hat{\Gamma}^4=\frac{1}{\sqrt{3}}(\hat{S}_x^2-\hat{S}_y^2),~~ \hat{\Gamma}^5=\hat{S}_{z}^2-\frac{1}{3}\hat{S}^2,
\end{equation}
which may be concisely written as
$
\hat{\Gamma}^a=\sum_{\alpha\beta}\hat{S}_\alpha Q_{\alpha\beta}^a\hat{S}_\beta.
$
Here, $Q^{a}$ are the five independent $3\times 3$ real  symmetric traceless matrices~\cite{Supp}. It is straightforward to check from the commutation relations $
[\hat{S}_{\alpha},\hat{S}_{\beta}]=\mathrm{i}\epsilon_{\alpha\beta\gamma} \hat{S}_\gamma$,
that $\hat{\Gamma}^a$ indeed generate $Cl_5$, namely, they satisfy the relations
\begin{equation}
\{\hat{\Gamma}^a,\hat{\Gamma}^b\}=2\delta^{ab}1_4
\end{equation}
and
\begin{equation}
\hat{\Gamma}^5=-\hat{\Gamma}^1\hat{\Gamma}^2\hat{\Gamma}^3\hat{\Gamma}^4.
\end{equation}

{\color{blue}\textit{The model}.}
As the Kitaev model is based on the Clifford algebra extension $Cl_{3}\cong Cl_{4}^{0}\subset Cl_{4}$, we now proceed to construct an exactly solvable model corresponding to $Cl_{5}\cong Cl_{6}^{0}\subset Cl_{6}$.

\begin{figure}
	\includegraphics[scale=1]{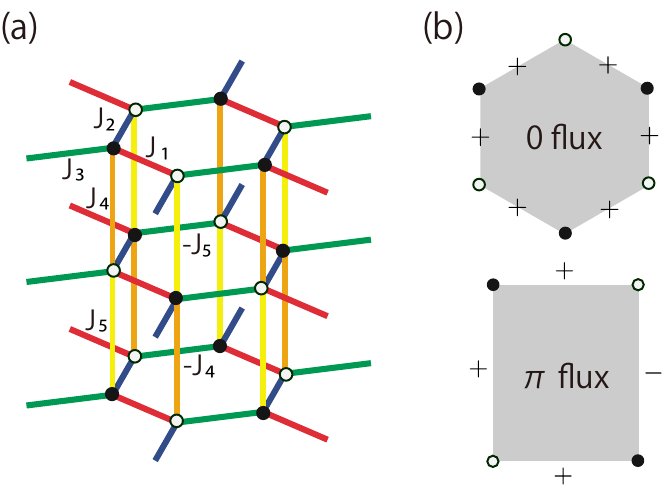}
	\caption{(a) Spin-3/2 model on a graphite lattice. The two sublattices are marked with black and white dots, respectively. The five bonds connected to a given site have different colors, corresponding to the five types of interactions $J_a$ with $a=1,2,\cdots,5$. Some vertical bonds have a negative sign, due to the ground-state gauge field configuration illustrated in (b), i.e., every hexagonal (rectangular) plaquette has flux $0$ ($\pi$). \label{fig:Stacked-Honeycomb}}
\end{figure}

Let us consider a spin-$3/2$ model on a graphite lattice, as illustrated in Fig.~\ref{fig:Stacked-Honeycomb}. The lattice is pentacoordinated, namely, each site is linked to five nearest neighbors: three within the ($x$-$y$) hexagonal layer and two along the vertical ($z$) direction. We label the five bonds by five different colors, corresponding to the five labels $a$ of $\hat{\Gamma}^a$ (see Fig.~\ref{fig:Stacked-Honeycomb} for a particular coloring scheme). Then our model can be expressed as
\begin{equation}\label{Exact-models}
\hat{H}=\sum_{\langle ij \rangle_a} J_a\hat{\Gamma}_i^a\hat{\Gamma}_j^a,
\end{equation}
where the summation is over all nearest neighbors. The interaction is quartic in terms of the spin operators [the explicit form can be obtained by substituting \eqref{Gamma-operators-1} and \eqref{Gamma-operators-2} into \eqref{Exact-models}] and is bond-dependent, i.e. the strength $J_a$ and the $\hat{\Gamma}^a$ involved both depend on the color of the bond. Like in the Kitaev model, such bond-dependent interactions introduce a strong exchange frustration, suppressing long range magnetic ordering.

To solve the model, we fractionalize the spin-3/2 at each site into six Majorana fermions $\gamma_i^\mu$ ($\mu=1,2,\cdots,6$).
As fermions, $\gamma^\mu$ at different sites anti-commute with each other. At each site $i$, they may be represented by $8\times 8$ matrices $\hat{\gamma}_i^\mu$, satisfying
\begin{equation}
\{\hat{\gamma}_i^\mu,\hat{\gamma}^\nu_j\}=2\delta_{ij}\delta^{\mu\nu}1_8.
\end{equation}
Clearly, this procedure doubles the Hilbert space at each site (from 4 to 8). The physical (even-parity) Hilbert space can be recovered by using the projection operator
\begin{eqnarray}\label{Projector}
\hat{\pi}_i=(\hat{p}_i+1_8)/2
\end{eqnarray}
for each site, where $\hat{p}_i=-\mathrm{i}\hat{\gamma}^1_i\hat{\gamma}^2_i\hat{\gamma}^3_i\hat{\gamma}^4_i\hat{\gamma}^5_i\hat{\gamma}^6_i$ is the fermion parity operator.

Then it is easy to check that in the physical subspace, $\hat{\Gamma}^a_{i}$ can be represented by
\begin{equation}\label{Maps}
\hat{\Gamma}^a_{i}\mapsto \mathrm{i}\hat{b}_i\hat{\gamma}^a_i,
\end{equation}
where $\hat{\gamma}^6_i$ has been distinguished as $\hat{b}_i$. It follows that our spin-$3/2$ models \eqref{Exact-models} can be mapped to a quartic Majorana model,
\begin{equation}\label{Majorana-models}
\tilde{H}=-\sum_{\langle ij\rangle_a}J_a\hat{u}_{\langle ij\rangle_a}\mathrm{i}\hat{b}_i\hat{b}_j,
\end{equation}
with $\hat{u}_{\langle ij\rangle_a}=\mathrm{i}\hat{\gamma}_i^a\hat{\gamma}_j^a$.


The solvability of our model \eqref{Majorana-models} relies on the key observation that all $\hat{u}$'s commute with each other and with all $\hat{b}$'s, and therefore commute with $\tilde{H}$, i.e., they are integrals of motion. Each $\hat{u}_{\langle ij\rangle_a}$ has eigenvalues $\pm 1$ on the bond, so it can be interpreted as a classical $\mathbb{Z}_2$ gauge field. Accordingly, the Majorana model \eqref{Majorana-models} can be interpreted as describing Majorana fermions ($\hat{b}$) hopping in the background of a $\mathbb{Z}_2$ gauge field ($\hat{u}$).

As numerically shown in \cite{Supp}, the gauge degree of freedom $\hat{u}$ in our model is gapped. Therefore, it is sufficient to focus on the ground-state flux sector. Since our model is defined on a bipartite lattice and is half-filled, by Lieb's theorem, the ground state must correspond to the gauge configuration with flux 0 for each hexagon plaquette and flux $\pi$ for each square plaquette (see Fig.~\ref{fig:Stacked-Honeycomb}), which is also confirmed by our numerical calculation~\cite{Supp}.
After fixing the gauge configuration, $\tilde{H}$ just becomes a noninteracting (quadratic) model of the Majorana fermion $\hat{b}$, and therefore can be easily solved~\cite{Note_Gauge_Equivalence}.

The above discussion demonstrates the exact solvability of our QSL model. It also becomes clear that as long as the pentacoordination is maintained, the solvability does not depend on the detailed lattice geometry nor the coloring scheme. The specific model in Fig.~\ref{fig:Stacked-Honeycomb} is one most natural realization. In the following, we will show that this model hosts intriguing topological features in the band structure of the emergent Majorana fermions, where the ground-state gauge configuration plays an essential role.

\begin{figure}
	\includegraphics[scale=0.55]{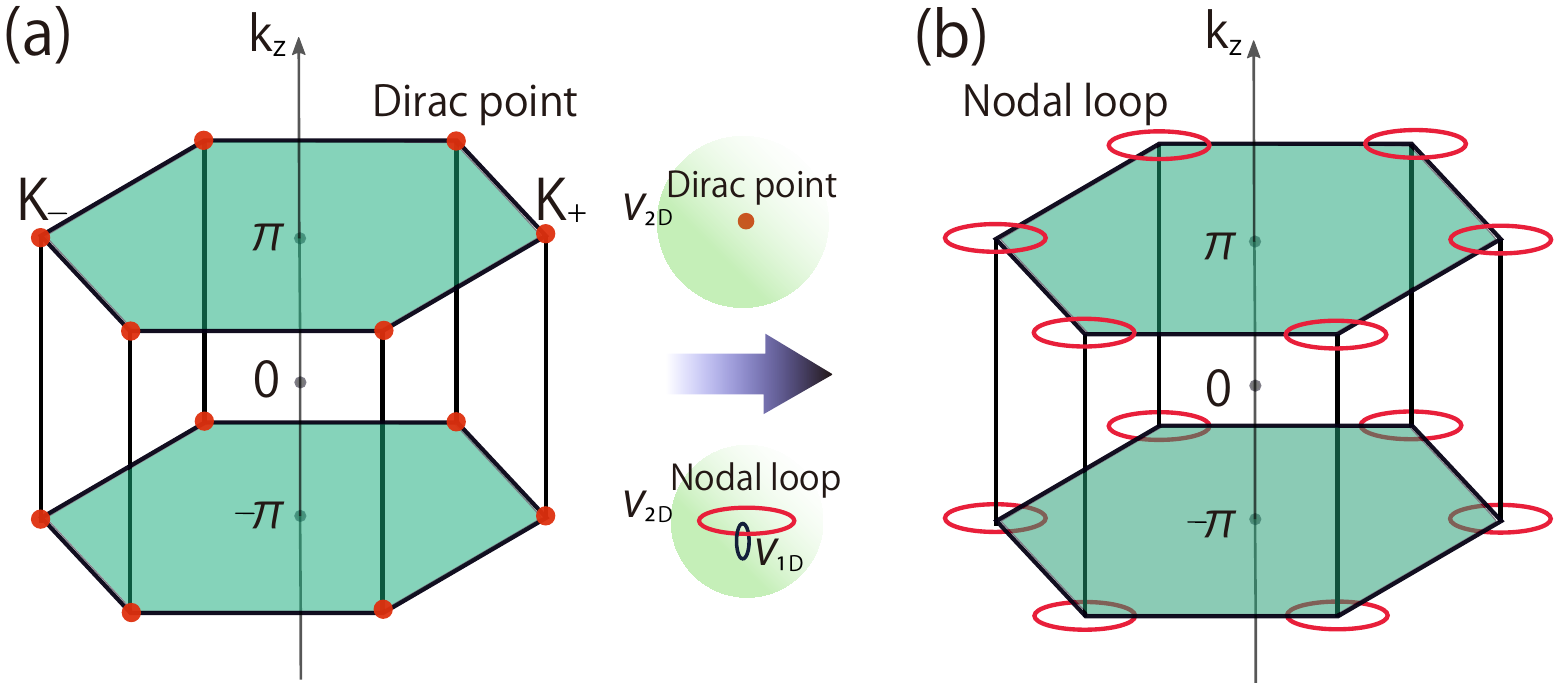}
	\caption{ (a) The critical state at $m=0$ is a 3D Dirac semimetal with real Dirac points located at the corners of the Brillouin zone. (b) When $m\neq 0$, each Dirac point is deformed into a nodal loop normal to $k_z$. The insets illustrate the topological charges $\nu_\text{2D}$ and $\nu_\text{1D}$. \label{fig:Hexagon-BZ}}
\end{figure}

{\color{blue}\textit{Topological charges of Majorana bands.}} Let's focus on the ground-state gauge configuration in Fig.~\ref{fig:Stacked-Honeycomb} and investigate its excitation spectrum.
Here, each unit cell of the Majorana model consists of four sites, so there are totally four energy bands. The momentum-space Hamiltonian can be derived as
\begin{equation}\label{Graphite-Model}
\mathcal{H}(\bm{k})=\sum_{a=1}^{3}f_a(\bm{k})\Gamma^a+g(k_z)~\mathrm{i}\Gamma^5\Gamma^3.
\end{equation}
The Hermitian Dirac matrices $\Gamma^a$ ($a=1,2,\cdots,5$) here are acting on the sublattice space, and should not be confused with the quadratic forms $\hat{\Gamma}^a$ of spin operators. Their explicit representations are given by $\Gamma^1=\sigma_0\otimes\tau_2$, $\Gamma^2=\sigma_3\otimes\tau_1$, $\Gamma^3=\sigma_2\otimes\tau_1$, $\Gamma^4=\sigma_1\otimes\tau_1$, and $\Gamma^5=\sigma_0\otimes\tau_3$,
with the $\sigma$'s and $\tau$'s the Pauli matrices and $\sigma_0$ the identity matrix. The $k$-dependent coefficients in (\ref{Graphite-Model}) are given by
\begin{equation}
  f_1(\bm{k})+\mathrm{i}f_2(\bm{k})=\sum_{i=1}^3J_i e^{\mathrm{i}\bm{k}\cdot\bm{b}_i},~ f_3(k_z)=J_{+}\cos (k_zc),
\end{equation}
and
\begin{equation}
  g(k_z)=J_{-}\sin (k_z c),
\end{equation}
where  $\bm{b}_i$ are the three in-plane bond vectors, $J_{\pm}=(J_5\pm J_4)/2$, and $c$ is the interlayer distance.

Before calculating the band structure, we point out an important symmetry of (\ref{Graphite-Model}): the off-centered spacetime inversion $\mathcal{P}T$ with $(\mathcal{P}T)^2=1$ (see \cite{Supp} for more details). First, in a $\Z_2$ gauge background, the Majorana fermions naturally preserve the time-reversal symmetry $T$, represented by $\hat{T}=\Gamma^5\hat{\mathcal{K}}\hat{I}$ for (\ref{Graphite-Model}), with $\hat{I}$ the momentum-inversion operator and $\hat{\mathcal{K}}$ the complex conjugation. Second, from Fig.~\ref{fig:Stacked-Honeycomb}, the model preserves the mirror reflection $M_z$ ($\hat{M}_z=\mathrm{i}\Gamma^4\Gamma^5\hat{I}_z$, with $\hat{I}_z$ the inversion of $k_z$),  meanwhile, it appears that the twofold screw rotation $\mathcal{S}_{2z}=\{C_{2z}|00\frac{1}{2}\}$ is violated, because the phases of the hopping amplitudes are not manifestly preserved. However, the invariance can be restored after a gauge transformation, so we do have the screw symmetry represented by $\hat{\mathcal{S}}_{2z}=\mathrm{i}\Gamma^4\Gamma^2\hat{I}_{xy}$, with $\hat{I}_{xy}$ the inversion of $k_x$ and $k_y$. The combination of $\mathcal{S}_{2z}$ and $M_z$ gives an \emph{off-centered} spatial inversion symmetry $\mathcal{P}$, represented by $\hat{\mathcal{P}}=\Gamma^2\Gamma^5\hat{I}$. Consequently, the model (\ref{Graphite-Model}) preserves the symmetry $\mathcal{P}T$ with $\hat{\mathcal{P}}\hat{T}=\Gamma^2\hat{\mathcal{K}}$. Importantly, this $\mathcal{P}T$ symmetry satisfies
$(\mathcal{P}T)^2=1$, which is in contrast to the usual $(PT)^2=-1$ expected for spin-3/2. This distinct symmetry property originates from the $\mathbb{Z}_2$ gauge configuration, and will strongly affect the topology of the model.


We now turn to the spectrum of the Majorana model \eqref{Graphite-Model}. In the following, we take $J_1=J_2=J_3=J$ for simplicity. First, at the critical point with $J_4=J_5$, the Fermi surface (at energy zero) consists of two isolated Dirac points $\bm{K}_{\pm}$ residing at the corners of the Brillouin zone (BZ) [Fig.~\ref{fig:Hexagon-BZ}(a)], making the spectrum resemble a 3D Dirac semimetal. Deviating from the critical point, i.e., when $m=J_5-J_4\neq 0$, the Dirac points are destroyed and each transformed into a nodal loop [Fig.~\ref{fig:Hexagon-BZ}(b)], so the generic phase of our model corresponds to a nodal-line semimetal with a pair of nodal loops.

When $m$ is sufficiently small compared with $J$, one can derive simple effective models expanded around $\bm{K}_{\pm}$. For example, around $\bm{K}_+$, we have
\begin{equation}\label{kp-model}
h(\bm{q})=\sum_{a=1}^3 v_a q_a\Gamma^a+ m~\mathrm{i}\Gamma^5\Gamma^3,
\end{equation}
where $\bm{q}=\bm{k}-\bm{K}_+$, $v_1=v_2=3Ja/2$, and $v_3=(J_4+J_5)c$. Clearly, for $m=0$, the model captures a Dirac point at $\bm q=0$, with fourfold degeneracy and linear dispersion. The nonzero $m$ term anti-commutes with the $q_3\Gamma^3$ term, such that it transforms the Dirac point into a nodal loop in the $x$-$y$ plane.


These nodal features carry and are protected by topological charges, which, in our model, are governed by the $\mathcal{P}T$ symmetry. 
Because $(\mathcal{P}T)^2=1$, the symmetry can always be put into the form $\hat{\mathcal{P}}\hat{T}=1_4\hat{\mathcal{K}}$ via a unitary transformation ($e^{\mathrm{i}\Gamma^2\pi/4}$ in our case).
It follows that the transformed Hamiltonian must be purely {real}.  The Dirac points at $m=0$ hence correspond to the \emph{real} Dirac points studied in Ref.~\cite{Zhao-PT-Dirac}. To characterize such real Dirac point, one can choose a sphere $S^2$ surrounding it in momentum space (see Fig.~\ref{fig:Hexagon-BZ}). According to Ref.~\cite{Zhao-Schnyder-Wang-PT}, the topological classification on this $S^2$ is given by
\begin{equation}\label{KO}
KO(S^2)\cong\mathbb{Z}_2,
\end{equation}
namely, the Dirac point carries a $\mathbb{Z}_2$ topological charge.
Specific for our four-band model in (\ref{Graphite-Model}), this charge can be evaluated by~\cite{Zhao-PT-Dirac}
\begin{equation}\label{Real-Chern}
\nu_\text{2D}=\frac{1}{4\pi}\int_{S^2}\mathrm{tr}(G\mathcal{F}_R) \mod 2,
\end{equation}
where $\mathcal{F}_R=d\mathcal{A}+\mathcal{A}\wedge\mathcal{A}$ is the real Berry curvature, $\mathcal{A}_{\alpha\beta}=\langle \alpha,\bm{k}|d|\beta,\bm{k}\rangle$ is the real Berry connection, with $|\alpha,\bm{k}\rangle$ and $|\beta,\bm{k}\rangle$ the real eigenstates of the two valence bands which correspond to the group $SO(2)$ [$\mathcal{F}$ and $\mathcal{A}$ are valued in the Lie algebra of $SO(2)$], and $G=-i\tau_2$ is just the $SO(2)$ generator~\cite{Note_Stable-equivalence}. This charge $\nu_\text{2D}$ is a special case of the so-called second Stiefel-Whitney number~\cite{Zhao-PT-Dirac,PhysRevB.96.155105,BJYang-Nodal-Line,B-J-Yang19APRPRX,wang2020boundary,Stiefel-Whitney,milnor2016characteristic}. This concept has been previously used in quantum field theory to judge whether a spin structure can be associated with a particular spacetime manifold~\cite{Stiefel-Whitney}. In both contexts, one needs to locally extend the real bundle to a spinor bundle, resulting in a sign ($\Z_2$) ambiguity for transition functions. This gives rise to a \v{C}ech cohomology class with $\Z_2$ coefficient, which corresponds to the second Stiefel-Whitney number.

When the $m$-term in Eq.~\eqref{kp-model} is smoothly tuned on, the Dirac point becomes unstable. However,
since $\mathcal{P}T$ is still maintained, the topological charge $\nu_\text{2D}$ on $S^2$ is unchanged, meaning that there are still band crossings inside $S^2$ (provided $m$ is sufficiently small). We already know this is the nodal loop, which hence inherits the topological charge $\nu_\text{2D}$. In addition, like ordinary nodal lines, the nodal loop here also have nontrivial Berry phases $\nu_\text{1D}$ over circles transversely enclosing it (see Fig.~\ref{fig:Hexagon-BZ}). Thus, the present $\mathcal{P}T$-invariant real nodal loops feature twofold topological charges $(\nu_\text{2D},\nu_\text{1D})\in \mathbb{Z}_2\times\mathbb{Z}_2$. It is well known that $\nu_\text{1D}$ leads to surface drumhead states bounded by the projection of each nodal line~\cite{PhysRevLett.113.046401,PhysRevB.92.045108}. As we shall see, $\nu_\text{2D}$ produces more interesting consequences on the boundary --- it determines a boundary criticality towards hinge Fermi arcs.

\begin{figure}
	\includegraphics[scale=1]{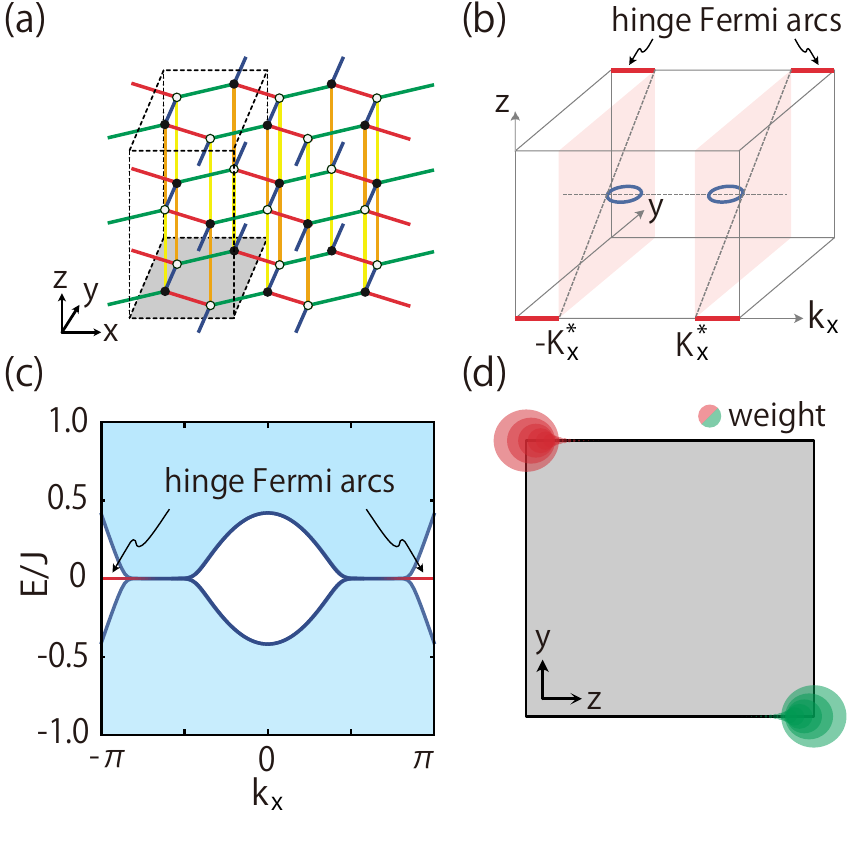}
	\caption{(a) The doubled unit cell for studying the boundary modes. (b) For a tube geometry, the second-order nodal-line semimetal phase has hinge Fermi arcs (marked in red) on a pair of hinges. (c) shows the corresponding energy spectrum. (d) shows the wave function distribution in the $y$-$z$ plane (i.e., cross-section of the tube) for the zero-modes at $k_x=\pi$. } \label{fig:Hinge-FermiArcs}
\end{figure}

{\color{blue}\textit{Hinge Fermi arcs.}} The hallmark of a topological state is the existence of topological boundary modes. We show that the generic phase of our model actually corresponds to a novel nodal-line semimetal with second-order topology, meaning that its topological zero-modes appear at second-order boundaries~\cite{PhysRevB.98.054432,PhysRevLett.110.046404,Benalcazar2017,PhysRevLett.119.246402,PhysRevLett.119.246401,Schindler2018a,PhysRevB.98.241103,PhysRevB.97.155305,PhysRevLett.123.186401,Wieder2020}, i.e., the hinges for a 3D system.

Let's first consider the critical state at $m=0$, i.e., the Dirac semimetal state. To investigate open boundaries, for convenience, we double the unit cell to have orthogonal in-plane lattice vectors [see Fig.~\ref{fig:Hinge-FermiArcs}(a)]. The two real Dirac points are then located on the $k_x$ axis at $\pm K_x$. The topological charge $\nu_\text{2D}$ lead to helical surface Fermi arcs on surfaces parallel to $x$, which connect the projections of the two Dirac points across the boundary ($k_x=\pi$) of the surface BZ (see \cite{Supp}). To understand this, one may view the 3D BZ as consisting of many 2D $k_y$-$k_z$ subsystems $H_{k_x}(k_y,k_z)$ parametrized by $k_x$, and each has a well-defined $\nu_\text{2D}$ as long as its spectrum is gapped. The locations $\pm K_x$ of the two Dirac points correspond to topological phase transition points
of these gapped 2D subsystems, such that $\nu_\text{2D}$ is nontrivial (trivial) outside (inside) the interval $[-K_x, K_x]$. Each nontrivial $H_{k_x}(k_y,k_z)$ gives a pair of helical states, which trace out the helical Fermi arcs on the side surfaces.

For the generic phase with $m\neq 0$, the bulk Dirac points are destroyed and the helical surface modes are no longer present~\cite{Supp}. However, since $\mathcal{P}T$ is respected, $\nu_\text{2D}$ for any gapped $H_{k_x}(k_y,k_z)$ is still well-defined and should remain the same as long as its gap does not close during the adiabatic turn-on of $m$. Hence, the 2D slices outside the interval $[-K_x^*, K_x^*]$ should retain a nontrivial $\nu_\text{2D}$, where $K_x^*$ is the maximum $k_x$ coordinate on the loop [see Fig.~\ref{fig:Hinge-FermiArcs}(b)]. The nontrivial bulk topological invariant means we must have topological zero-modes somewhere on the boundary. Now, as the surface is ruled out, where are these zero-modes located?

The answer is at the hinge. In Fig.~\ref{fig:Hinge-FermiArcs}(c), we show the numerical result for a tube geometry which is extended along $x$ but confined in $y$ and $z$. In the spectrum, one indeed observes zero-modes in the interval $[-\pi,-K_x^*]\cup[K_x^*,\pi)$. By checking their wave function distribution [Fig.~\ref{fig:Hinge-FermiArcs}(d)], we confirm that these modes are located at the hinges. Interestingly, these hinge Fermi arcs exist only at a single ($\mathcal{P}T$-connected) pair of the hinges between the side surfaces; whereas the other pair of the hinges are gapped [Fig.~\ref{fig:Hinge-FermiArcs}(b, d)]. Which pair to occupy is determined by the sign of $m$. The two second-order nodal-line semimetal phases with $m>0$ and $m<0$ are thus distinguished by the location of their hinge Fermi arcs, and they are separated by the critical state of the Dirac semimetal at $m=0$.

{\color{blue}\textit{Discussion.}}
We have proposed a new exactly solvable model with QSL ground state beyond the Kitaev model. Remarkably, this model also yields novel second-order topological features not seen before.

As mentioned, the model should remain exactly solvable when we change the lattice geometry or the coloring scheme but maintain the coordination. This offers vast opportunities to explore interesting physics in our QSL model. For example, in the Supplemental Material~\cite{Supp}, we also consider another coloring scheme, which differs from Fig.~\ref{fig:Stacked-Honeycomb} in the arrangement of the vertical bonds. It has the same ground-state gauge configuration, but entirely different physics. Particularly, the Majorana fermions have the $PT$ symmetry with $(PT)^2=-1$, leading to the twofold Kramers degeneracy for each band. The resulting state corresponds to a topological insulator phase with surface Dirac points.


\begin{acknowledgements}
{\color{blue}\textit{Acknowledgments.}} We thank Q.-H. Wang for helpful discussions and thank S. Li for help with the figures. This work is supported by the NSFC (Grant No.~11874201), the Fundamental Research Funds for the Central Universities (Grant No.~0204/14380119), and the Singapore MOE AcRF Tier 2 (MOE2019-T2-1-001).
\end{acknowledgements}
\bibliographystyle{apsrev}
\bibliography{Second-order_Spin_Liquids}

\begin{thebibliography}{40}
\bibitem[{Not({\natexlab{a}})}]{Note_Gauge_Equivalence}
\bibinfo{note}{For gauge equivalent configurations, $\tilde{H}$ has the same
  eigenenergy, and it is easy to show that their eigenstates are projected into
  the same physical state by the aforementioned projector $\hat{\Pi}=\prod_i
  \hat{\pi}_i$.}
\bibitem[{Not({\natexlab{b}})}]{Note_Stable-equivalence}
\bibinfo{note}{The $K$-theoretical classification requires stable equivalence,
  which means an arbitrary number of trivial bands can be added to the
  concerned band theory. Therefore, $SO(2)$ is considered as a subgroup of a
  larger $SO(N)$, and only the parity of the fundamental group of $SO(2)$ is a
  stable topological invariant.}
\end{thebibliography}

\clearpage
\newpage

\onecolumngrid
\appendix

\section{Supplemental Materials for ``Topological second-order spin-$\frac{3}{2}$ liquids with hinge Fermi arcs"}


\section{Symmetric and traceless quadratic forms}
The five independent symmetric and traceless $3\times 3$ quadratic forms are explicitly given by
\begin{equation}
Q^1=\frac{1}{\sqrt{3}}\begin{pmatrix}
0 & 0 & 0\\
0 & 0 & 1\\
0 & 1 & 0
\end{pmatrix},~~Q^2=\frac{1}{\sqrt{3}}\begin{pmatrix}
0 & 0 & 1\\
0 & 0 & 0\\
1 & 0 & 0
\end{pmatrix},~~Q^3=\frac{1}{\sqrt{3}}\begin{pmatrix}
0 & 1 & 0\\
1 & 0 & 0\\
0 & 0 & 0
\end{pmatrix},
\end{equation}
and
\begin{equation}
Q^4=\frac{1}{\sqrt{3}}\begin{pmatrix}
1 & 0 & 0\\
0 & -1 & 0\\
0 & 0 & 0
\end{pmatrix},~~Q^5=\frac{1}{3}\begin{pmatrix}
-1 & 0 & 0\\
0 & -1 & 0\\
0 & 0 & 2
\end{pmatrix}.
\end{equation}

\section{The ground-state flux configuration and static vortex mass}

We numerically verify the ground-state flux configuration on the graphite lattice of size $(L \mathbf{n}_1, L \mathbf{n}_2+\mathbf{n}_1, L_z \mathbf{n}_3 )$ with periodic boundary condition, where the basis vectors read $\mathbf{n}_1 = \{\frac{1}{2}, \frac{\sqrt{3}}{2}, 0\}$, $\mathbf{n}_2 = \{-\frac{1}{2}, \frac{\sqrt{3}}{2}, 0\}$, and $\mathbf{n}_3 = \{0, 0, 1\}$. By setting $J_a=1$ for $a=1$, 2, 3 and $J_4=J_5=0$, we recover Kitaev's honeycomb model with ground state energy extrapolating to $E_0 \approx -1.5746$ per unit cell (u.c.). The single vortex energy $\Delta E_\mathrm{vortex} = E_\mathrm{vortex}(L)-E_0(L)$ extrapolates to $\Delta E_\mathrm{vortex} \approx 0.1536$. This is in agreement with Kitaev's results.

\begin{figure}
	\includegraphics[scale=0.6]{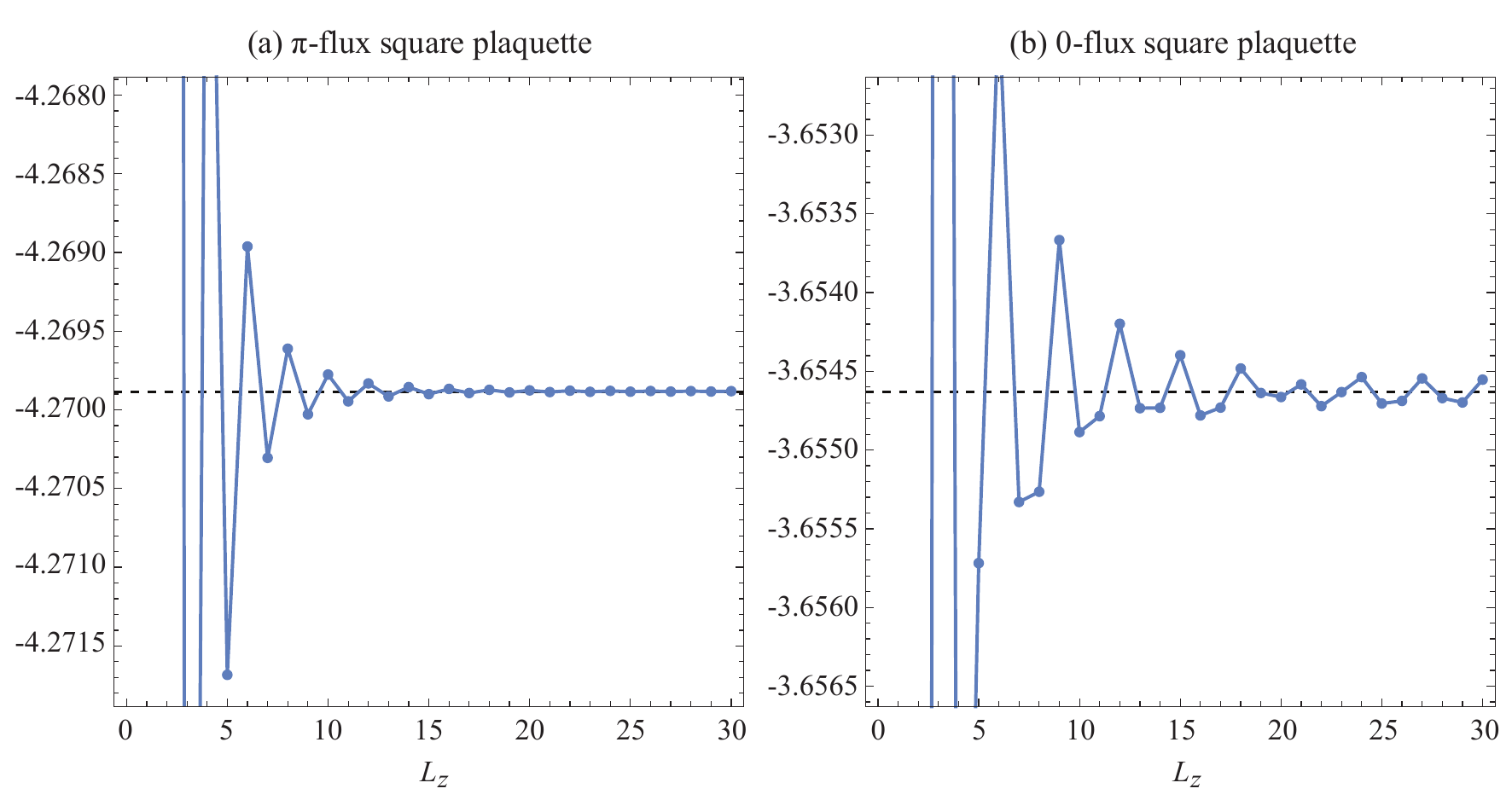}
	\caption{\label{fig:0vsPi} Ground-state energy per unit cell for flux configurations with $0$ flux in each hexagonal plaquette and (a) $0$ flux or (b) $\pi$ flux in each square plaquette.}
\end{figure}

In the following, we set $J_4=J_5=1$, which renders the two configurations in Fig.~1 equivalent. We note that other values of $J_4$ and $J_5$ are also tested, and qualitatively the same results are found. We further fix $L=32$, which gives relatively well-converged results for both the ground-state and single-vortex energies in the 2D case. Figure~\ref{fig:0vsPi}(a) and (b) show the ground-state energy per u.c. for flux configurations with $0$ flux in each hexagonal plaquette and $\pi$ flux or $0$ flux in each square plaquette, respectively. The latter can be realized by setting the negative interlayer coupling terms $-J_{4/5}$ to $J_{4/5}$. With increasing $L_z$, the former configuration converges to $E_0 \approx -4.2699$ and the latter to $-3.6546$.

\begin{figure}
	\includegraphics[scale=0.6]{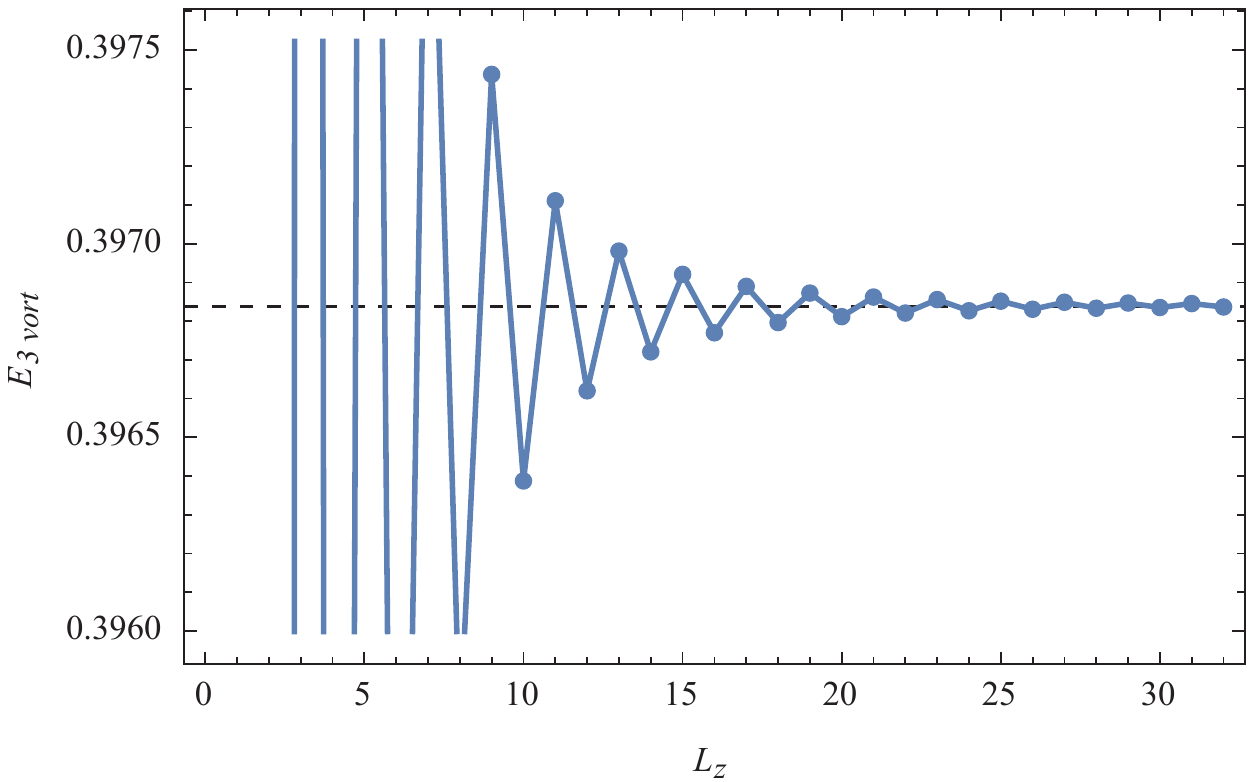}
	\caption{\label{fig:3vortices} The energy of three edge-sharing vortices on the square plaquettes.}
\end{figure}

Now we can further test the validity of the ground state by creating vortices in the $\pi$-flux phase. In the 2D case, string operators can be defined to create isolated vortices. Here, however, for a given unit cell, reversing either one of its in-plane or out-of-plane bonds changes the flux of its two square plaquettes at the same time. Therefore, it is not possible to create isolated vortices (one flux change in a u.c.) as in the 2D case. For simplicity, we create three edge-sharing vortices on the square plaquettes by reversing one single vertical bond. Figure~\ref{fig:3vortices} shows three-vortex energy as a function of $L_z$ (with fixed $L=32$), which converges towards $E_\mathrm{3vort} \approx 0.3972$.

\section{Surface States}
\begin{figure}
	\includegraphics[scale=1]{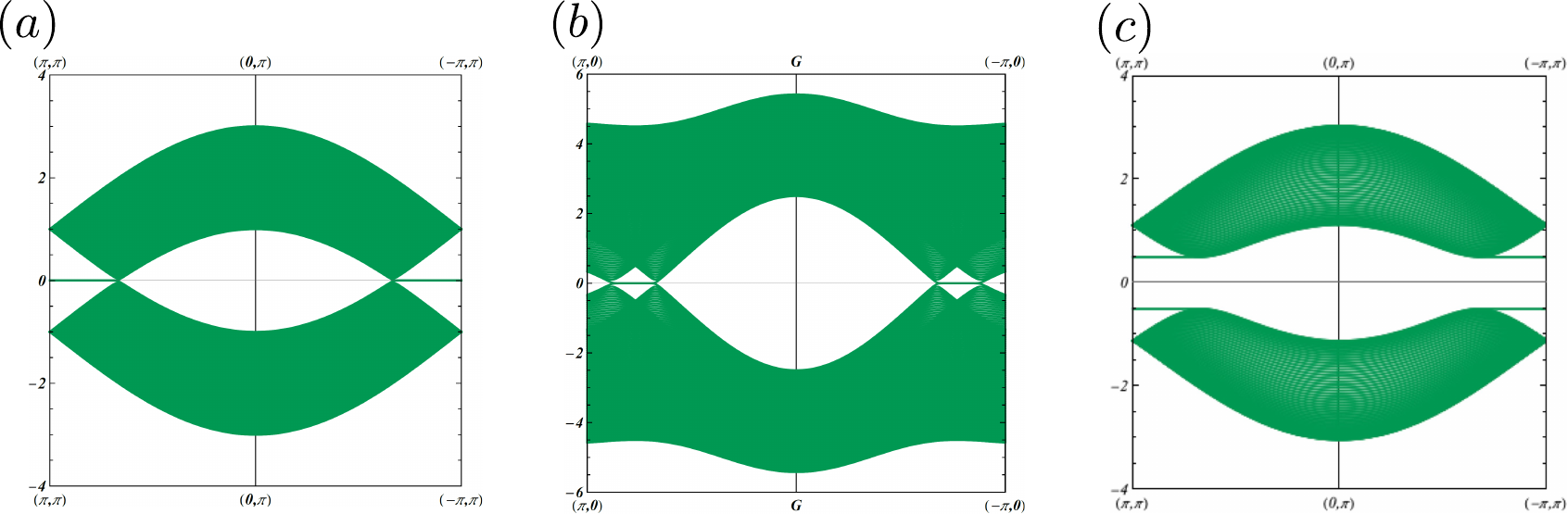}
	\caption{Surface states. (a) The Fermi arcs on the zigzag-$z$ surface in the real Dirac semimetal phase. (b) The drumhead states on the $xy$-surface in the nodal-line semimetal phase when $J_{-}\ne 0$. (c) The gapped spectrum for the zigzag-$z$ surface in the nodal-line semimetal phase. \label{Surface-States}}
\end{figure}

In the Dirac semimetal phase, helical Fermi arcs appear on the surfaces parallel the the zigzag direction, as shown in  Fig.\ref{Surface-States}(a). The Fermi arcs connect the projections of the two Dirac points crossing the boundary of the Brillouin zone. In the nodal-line phase, drumhead states appear on the surface normal to the $z$-direction [Fig.\ref{Surface-States}(b)], and are bounded by the projected images of the nodal lines in the surface Brillouin zone. The drumhead states come from the $1$D topological charges of the nodal lines. In the nodal-line semimetal phase, the zigzag-$z$ surface spectrum has a finite energy gap as shown in Fig. \ref{Surface-States}(c).


\section{Space symmetries of the model in the main text}
In this section we discuss the representation of space symmetries in our model in detail. The translational distance for the unit cells along the $z$-direction is $2c$. Hence, the Hamiltonian in the main text is not periodic in the first Brillouin zone for $k_z\in[-\pi/2c,\pi/2c)$. For the non-periodic Hamiltonian, the twofold screw rotation symmetry $\mathcal{S}_{2z} $ and mirror reflection symmetry $M_z$ are represented by
\begin{equation}
\begin{split}
\hat{\mathcal{S}}_{2z} &=\sigma_2\otimes\tau_0 \hat{I}_{xy}=\mathrm{i}\Gamma^4\Gamma^2\hat{I}_{xy},\\
\hat{M}_z &=\sigma_1\otimes\tau_2\hat{I}_z=\mathrm{i}\Gamma^4\Gamma^5\hat{I}_z.
\end{split}
\end{equation}
As we see from Fig.1(a) in the main text, the twofold screw rotation symmetry $\mathcal{S}_{2z}$ is nonsymmorphic. In other words, it is the twofold rotation followed by a half translation along the $z$-direction.
The combination of $\mathcal{S}_{2z}$ and $M_z$ is the off-centered spatial inversion symmetry $\mathcal{P}$. $\mathcal{P}$ and time reversal symmetry $T$ are represented by
\begin{equation}
\begin{split}
\hat{\mathcal{P}} &=\hat{\mathcal{S}}_{2z}\hat{M}_z=-\mathrm{i}\sigma_3\otimes\tau_2\hat{I}=\Gamma^2\Gamma^5\hat{I},\\
\hat{T} &=\sigma_0\otimes\tau_3\hat{\mathcal{K}}\hat{I}=\Gamma^5\hat{\mathcal{K}}\hat{I}.
\end{split}
\end{equation}
The combination of time-reversal and the off-centered inversion is the off-centered spacetime inversion symmetry $\mathcal{P}T$, which is represented as
\begin{equation}
\begin{split}
\hat{\mathcal{P}}\hat{T}=\sigma_3\otimes\tau_1=\Gamma^2\hat{I}.
\end{split}
\end{equation}

To see the effects of half-translation in $\mathcal{S}$ for the representations of relevant symmetry operators, we need  to resume the periodicity of the Hamiltonian. This is implemented by the unitary transformation
\begin{equation}
V(k_z)=\begin{pmatrix}
e^{\mathrm{i} k_z/4} & 0 \\
0 & e^{-\mathrm{i} k_z/4}
\end{pmatrix}\otimes\tau_0.
\end{equation}
Then, the interlayer terms are transformed as
\begin{equation}
\begin{split}
& V(k_z)\left(2J_+\cos \frac{k_z}{2} \sigma_2\otimes\tau_2+2J_-\sin \frac{k_z}{2}\sigma_2\otimes\tau_2\right) V^\dagger(k_z)\\
=& J_+(1+\cos k_z)\sigma_2\otimes\tau_1+J_+\sin k_z\sigma_1\otimes\tau_1 + J_{-}(1-\cos k_z)\sigma_1\otimes\tau_2+J_{-}\sin k_z \sigma_2\otimes\tau_2\\
=& J_{+}[(1+\cos k_z)\Gamma^3+\sin k_z \Gamma^4]+J_{-}[(1-\cos k_z)\mathrm{i}\Gamma^4\Gamma^5+\sin k_z\mathrm{i}\Gamma^3\Gamma^5],
\end{split}
\end{equation}
where
\begin{equation}
J_{\pm}=\frac{1}{2}(J_5\pm J_4).
\end{equation}
It is manifest that the periodicity along $k_z$ is satisfied.

The twofold screw rotation operator $\hat{\mathcal{S}}_{2z}$ is accordingly transformed as
\begin{equation}
\begin{split}
\hat{\mathcal{S}}'_{2z}=V(k_z)\hat{\mathcal{S}}V^\dagger(k_z) &=\begin{pmatrix}
0 & -\mathrm{i} e^{\mathrm{i} k_z/2} \\
\mathrm{i} e^{-\mathrm{i} k_z/2} & 0
\end{pmatrix}\otimes\tau_0 \hat{I}_{xy}\\ &=-\mathrm{i}
\mathcal{G}\begin{pmatrix}
0 & e^{\mathrm{i} k_z/2}\\
e^{-\mathrm{i} k_z/2} & 0
\end{pmatrix}\otimes\tau_0 \hat{I}_{xy}
\end{split}
\end{equation}
with
\begin{equation}\label{Gauge}
\mathcal{G}=\sigma_3\otimes\tau_0.
\end{equation}
Now it is manifest that the momentum dependence of the operators comes from the half translation in the screw rotation. The followed $\mathcal{G}$ is the $\Z_2$ gauge transformation performed to restore the $\Z_2$ phases in the  original hopping pattern in Fig.1(a) in the main text.

$M_z$ and $T$ are symmorphic, and therefore their representations are invariant under the transformation $V(k_z)$. Hence, the operators $\hat{\mathcal{P}}$ and $\hat{\mathcal{P}}\hat{T}$ are transformed as
\begin{equation}
\hat{\mathcal{P}}'=V(k_z)\hat{\mathcal{P}}V^\dagger(k_z)=\begin{pmatrix}
-\mathrm{i} e^{\mathrm{i} k_z/2} & 0\\
0 & \mathrm{i} e^{-\mathrm{i} k_z/2}
\end{pmatrix}\otimes\tau_2\hat{I},
\end{equation}
and
\begin{equation}
\hat{\mathcal{P}}'\hat{T}=V(k_z)\hat{\mathcal{P}}\hat{T}V^\dagger(k_z)=\begin{pmatrix}
e^{\mathrm{i} k_z/2} & 0\\
0 & -e^{-\mathrm{i} k_z/2}
\end{pmatrix}\otimes\tau_1 \hat{\mathcal{K}}.
\end{equation}
The momentum-dependence for the operators comes from the fact that they are off-centered.

\section{Another exactly solvable model and its symmetries}
\begin{figure}
	\includegraphics[scale=1]{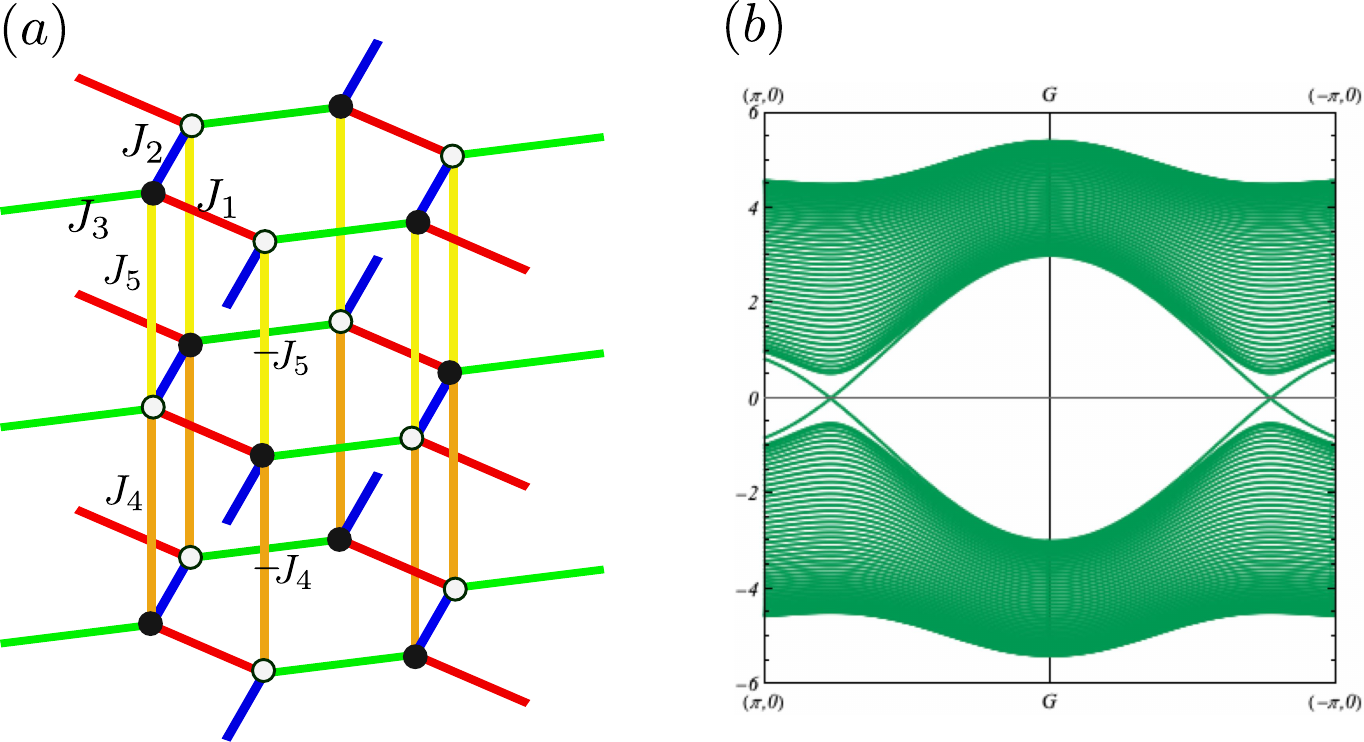}
	\caption{\label{fig:Lattice-2} The coloring of the exactly solvable model and the Dirac points related by mirror symmetry on the zigzag-$z$ surface in the crystalline topological superconductor phase.}
\end{figure}
The second exactly solvable model is illustrated in Fig.\ref{fig:Lattice-2}. The tight-binding model for the Majorana spinors is given by
\begin{equation}\label{Graphite-Model}
\mathcal{H}^c(\bm{k})=\sum_{a=1}^{3}f_a(\bm{k})\Gamma^a+g(k_z)\Gamma^4.
\end{equation}
The last term violates the twofold screw rotation symmetry $\mathcal{S}_{2z}$, but preserves $\hat{M}_z$ and $\hat{T}$. This can also be seen clearly from the hopping pattern in Fig.\ref{fig:Lattice-2}. Since $PT$ symmetry is broken, the Dirac points lost their protective symmetry, and the spectrum is fully gapped. Actually, the Majorana spinors are in a crystalline topological superconductor phase in class BDI. Two Dirac points reside on any surface parallel to the zigzag direction, which are related by the mirror symmetry $M_x$. 

We now discuss the symmetries of the model. It has the symmorphic twofold rotation symmetry $C_{2z}$, which is represented as
\begin{equation}
\hat{C}_{2z}=\mathcal{G}\,\sigma_0\otimes \tau_2\hat{I}_{xy}=\mathrm{i}\Gamma^2\Gamma^5 \hat{I}_{xy},
\end{equation}
where the spatial rotation $\sigma_0\otimes\tau_2$ is followed by the gauge transformation $\mathcal{G}$, Eq.~\eqref{Gauge}, to restore the phases in the hopping pattern. The mirror symmetry $M_z$ is represented by
\begin{equation}
\hat{M}_z=\sigma_1\otimes\tau_2\hat{I}_z=\mathrm{i}\Gamma^4\Gamma^5\hat{I}_z.
\end{equation}
We observe that the gauge transformation $\mathcal{G}$ anti-commutes with $\hat{M}_z$,
\begin{equation}\label{Modified-commu}
\{\mathcal{G},\hat{M}_z\}=0.
\end{equation}
Because of this, the commutation relation of $M_z$ and $C_{2z}$ is projectively modified by the $\Z_2$ coefficient as
\begin{equation}
\{\hat{M}_z,\hat{C}_{2z}\}=0.
\end{equation}
The (centered) spatial inversion symmetry is represented by the combination,
\begin{equation}
\hat{P}=\hat{C}_{2z}\hat{M}_z\hat{I}=\Gamma^2\Gamma^4\hat{I}.
\end{equation}
Because of Eq.~\eqref{Modified-commu}, we find
\begin{equation}
\hat{P}^2=-1.
\end{equation}
The time-reversal symmetry $T$ is still represented by $\hat{T}=\Gamma^5\hat{\mathcal{K}}\hat{I}$, 
which commutes with $\hat{P}$,
\begin{equation}
[\hat{P},\hat{T}]=0.
\end{equation}
The spacetime inversion symmetry $PT$ is represented by
\begin{equation}
\hat{P}\hat{T}=\i\sigma_2\otimes\tau_3\hat{\mathcal{K}}=\Gamma^3\Gamma^2\hat{\mathcal{K}}.
\end{equation}
It satisfies
\begin{equation}
(\hat{P}\hat{T})^2=-1
\end{equation}
as claimed in the main text. Hence, the (centered) spacetime inversion is consistent with that of the spin-$\frac{3}{2}$, whose square is also equal to $-1$.

\end{document}